# *VCEMO:* Multi-Modal Emotion Recognition for Chinese Voiceprints

Jinghua Tang[1, 3, *], Liyun Zhang[1, 2, *], Yu Lu[1, 2], Dian Ding[1, 2, †], Lanqing Yang[1, 2], Yi-Chao Chen[1, 2], Minjie Bian[2, 4], Xiaoshan Li[2, 4, †], Guangtao Xue[1, 2]

[1] Shanghai Jiao Tong University, Shanghai, China.
[2] Shanghai Key Laboratory of Trusted Data Circulation and Governance in Web3.
[3] Shanghai Voicecomm Information Technology Co., Ltd., Shanghai, China
[4] Shanghai Data Group Co., Ltd, Shanghai, China 200011
`dingdian94@sjtu.edu.cn; lixiaoshan@shdatagroup.com`

**Abstract.** Emotion recognition can enhance humanized machine responses to user commands, while voiceprint-based perception systems can be easily integrated into commonly used devices like smartphones and stereos. Despite having the largest number of speakers, there's a noticeable absence of high-quality corpus datasets for emotion recognition using Chinese voiceprints. Hence, this paper introduces the *VCEMO* dataset to address this deficiency. The proposed dataset is constructed from everyday conversations and comprises over 100 users and 7,747 textual samples. Furthermore, this paper proposes a multimodal-based model as a benchmark, which effectively fuses speech, text, and external knowledge using a co-attention structure. The system employs contrastive learning-based regulation for the uneven distribution of the dataset and the diversity of emotional expressions. The experiments demonstrate the significant improvement of the proposed model over SOTA on the *VCEMO* and *IEMOCAP* datasets. *Code and dataset will be released for research.*

## 1 Introduction

Audio data plays a fundamental and irreplaceable role in our comprehension of the world. It encapsulates not only words and language but also the intricate tapestry of human experiences and emotions. Consequently, audio data exhibits an astonishingly diverse array of applications that span the entire spectrum of human endeavor. Specifically, voiceprint-based emotion recognition from audio data is paramount for assistance in communicating with people and many human-computer interaction applications. In call centers, employees can make informed decisions by receiving timely feed-back on customers' moods or assess business interactions based on customers' emotional states. Simultaneously, the software application can adapt and enhance user experiences by implementing appropriate behaviors through real-time monitoring of the user's emotions [24].

---

[*] Both authors contributed equally to the research.
[†] Dian Ding and Xiaoshan Li are the corresponding authors.

With over a billion native speakers and a rich cultural heritage, Chinese has undeniably emerged as a highly popular language on the global stage. Hence, the application of emotion recognition for Chinese audio holds great promise. However, there exists a restricted amount of Chinese corpus data available for model training in the context of emotion recognition. Furthermore, current methods are typically trained and evaluated on English datasets, lacking specific processing and optimization for Chinese data.

Consequently, we collected a large Chinese conversation sentiment corpus called VCEMO for the single-sentence Chinese emotion recognition task. The dataset consists of single-sentence conversations of everyday life and has several advantages:

**1). Rich voiceprint information**: Considering that the collection of voice information in previous datasets (e.g., CASIA [18], *IEMOCAP* [1]) has often relied on a specific few professional readers or professional actors, only a few people's pronunciation information as well as voiceprint features are present in the datasets. Our dataset contains daily speech data from more than 100 people, including a wide range of Chinese pronunciation accents and spoken language features.

**2). Abundant text information**: The textual content of the dataset is exclusively sourced from spontaneous conversations in everyday life. Consequently, there exist substantial disparities between these texts, and they are abundant in information.

**3). Adaptability to multi-modal fusion**: Given that the data originate exclusively from everyday conversations and individuals naturally employ various textual expressions to convey their inner sentiments based on their emotions, we can effectively leverage the multi-modal fusion of audio signals and textual information for the emotion recognition task.

Contemporary methods [5, 4, 15, 20, 21,19] commonly employ neural networks for tasks such as emotion recognition, as well as for effective feature extraction and classification of data. Given the notable distinctions between Chinese and English, these methods lack specific processing tailored to Chinese information. Hence, leveraging the extensive Chinese corpus dataset *VCEMO*, we introduce a novel multimodal model for emotion recognition. Automatic speech recognition (ASR) [11] is adapted to convert audio signals into Chinese text messages. For Chinese text, we use the pre-trained Chinese BERT architecture for processing. In addition, we utilize text embedding for additional emotion feature extraction from Chinese text. Finally, the co-attention structure is employed to fuse multi-modal data features.

Furthermore, given that the *VCEMO* dataset originates from everyday conversations, there is an uneven distribution of emotional data within the dataset. Additionally, a single audio sentence may encompass diverse emotional expressions, making it challenging for a singular emotional label to fully convey its comprehensive emotional information. To tackle the aforementioned issue, we employ a contrastive-learning-based regulation for training our model. Eventually, experimental tests have demonstrated that our model has significantly better emotion recognition performance on *VCEMO* than previously studied models.

Overall, our contributions are as follows:

– We produce a new Chinese daily conversational corpus dataset for emotion recognition, called *VCEMO*, containing 7477 samples of audio signals from over 100 individuals.

– We propose a multi-modal model for acoustic data and text data (word embeddings and pre-trained BERT embeddings) using the co-attention structure for multi-modal feature fusion.

– We employ a contrastive-learning-based regulation to train and optimize models, mitigating issues related to sample imbalance and under-representation of individual labels.

– Extensive experiments show that our model has SOTA performance on the *VCEMO* and *IEMOCAP* datasets.

## 2  Related Work

Speech emotion recognition has been studied for multiple decades within both the machine learning and speech communities. In alignment with the prevailing research approach, scholars extract feature insights from audio data and subsequently employ these insights across a range of classifiers, including: hidden Markov models [12], convolutional recurrent network [16], SVM [13], hierarchical binary decision tree [8], gaussian mixture [3], nerual network [14]. Much of the aforementioned works relied on context to furnish additional information for correcting and inferring emotional content extracted from the data. The mining and analysis of emotional information from single-sentence audio data can pose more significant challenges. Xu et al. [20] introduced an attention-based network designed for aligning textual and audio information, along with feature extraction. Yoon [21, 22] presented a groundbreaking deep dual recurrent encoder model that seamlessly merges text data and audio signals. This model employs a pair of recurrent neural networks (RNNs) to holistically encode the information. Delbrouck [2] et al. proposed a transformer-based joint-encoding model called UMNOS for single-sentence emotion recognition and sentiment analysis.

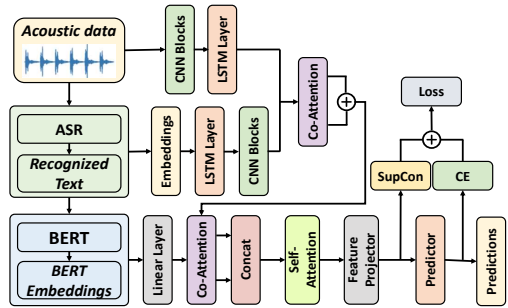

**Fig. 1.** The multi-modal model for emotion recognition.

## 3      Our Approach

In this section, we describe our emotion recognition model. This model employs three distinct modalities of data as input sources: acoustic signals, word embeddings, and BERT-encoded embeddings. Initially, each modality is processed separately. Subsequently, all the features from the various input modalities are combined using a co-attention layer. Finally, Linear layers are employed to produce the predictions. The overall model structure is shown in Fig. 1.

### 3.1      Modality Input

This model has three modalities as input. Regarding the acoustic input, we utilized the mel-spectrogram, which is generated by applying a Short-Time Fourier Transform (STFT) to the audio signal. The mel-spectrogram provides a visual representation of the energy in different frequency bands of an audio signal changing over time, with the frequency axis adjusted to better match the human auditory perception.

ASR is a technology that converts audio data into text data, facilitating the transcription and understanding of spoken words by machines. We use the ASR module to extract recognized text from audio signals. For text input, we use text embedding to learn text features directly. Meanwhile, we incorporate pre-trained BERT to extract transcription features from external knowledge.

### 3.2      Modality Pre-process

After retrieving the mel-spectrogram of the audio signals, we apply a classic Conv-BatchNorm-ReLU structure to extract features in both the time and frequency dimensions. Then, an LSTM layer is applied to extract deeper features in the time dimension. Additionally, the word embeddings have a better time structure and are more straightforward in each time slot. Hence, an LSTM is applied to the word embeddings before using a 1D-convolution layer to incorporate the information from the entire timeline. The feature extracted from BERT is a 768-dimensional vector. As it is already well-structured and contains abundant information, we applied a Linear layer to modify its size for subsequent multi-modal fusion and information compression.

### 3.3      Multi-modal fusion

Given the presence of three modalities, we need two rounds of fusion to comprehensively combine all the information extracted from these different modalities, and determining the order of fusion is a significant consideration. In our model, we first fuse the audio features and word embedding features. Their akin temporal structures make them suitable for initial fusion, as this process enhances the temporal dimension by leveraging their shared characteristics to amplify common information and compensate for missing data unique to one modality. Subsequently, the time-structured feature mentioned earlier is fused with the BERT-encoded feature, incorporating external knowledge from the outside world to in-dataset knowledge. In each fusion, there are two stages: extracting additional features from one modality with knowledge from another modality and then merging these additionally extracted features into a single representation.

In the first stage, we employed the co-attention layer to convey the presence of another modality to each modality. The structure of co-attention layer is as shown in Fig. 2. Inspired by [23], we employed the Encoder-Decoder structure to stack multiple layers of attention modules. In the co-attention layer, the first modality employs self-attention alone to extract deeper information from itself. Following that, the second modality goes through a self-attention operation, during which a guided-attention step is conducted to extract more information while considering both modalities. In contrast to simply using the output of the self-attention from another modality at the same depth as the input for guidedattention, leveraging the final output of the Self-attention layers can offer more enriched information and a more accurate guide. Both self-attention and guidedattention are based on the attention mechanism [17]. The attention module aids in constructing a holistic perspective of the entire period during the speech. The attention consists of a query q, a key k and a value v:

$$Attention(q, k, v) = softmax(\frac{qk^T}{\sqrt{d}})v$$

In the self-attention, all of q, k, and v are from the same modality. However, in guided-attention, the v and k are from the same modality while q is from another modality.

The first stage of the two fusion is the same, yet they diverge in the second stage. Considering the similarity of time structures, for the fusion between features from audio data and word embeddings, we employ a straightforward element-wise addition. This approach enhances their temporal structure and reduces the feature size compared to concatenation. In the second fusion, the features are dissimilar and lack a shared temporal structure, which leads to lossy and disorganized information when using element-wise addition. Consequently, concatenation is employed to retain more information, which is crucial for effectively leveraging both in-dataset knowledge and external-world knowledge. Following the ultimate fusion, we applied additional self-attention to comprehensively process the collective information from all modalities and proceed to make predictions using a two-layer MLP.

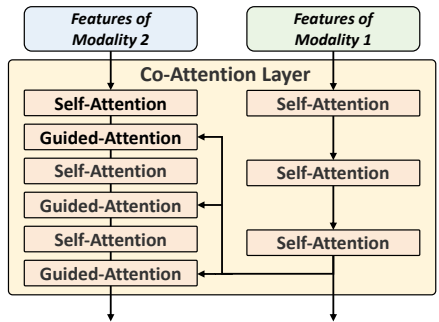

**Fig. 2.** The architecture of the co-attention layer

### 3.4 Contrastive Learning

Through our examination of misclassified cases in current state-of-the-art models, we identified that the ambiguity in the emotions expressed by actors is another factor hindering the model from learning accurate features. It is common to observe that a person's emotions can be complex, even involving contradictory feelings simultaneously. However, datasets with labels assigned to a single emotion as the ground truth may be misleading in capturing the presence of other coexisting emotions. Furthermore, employing traditional cross-entropy loss during model training mechanically steers the model to predict a probability of 1 only for the labeled emotion, penalizing predictions with non-zero probabilities for other emotions. This situation can significantly perplex the model, especially in cases where multiple emotions coexist. Moreover, stemming from naturalistic conversations in daily life, our dataset exhibits an imbalanced distribution of labels. Specifically, there is a pronounced prevalence of sentences labeled as neutral, contrasting with a scarcity of instances labeled as surprise.

Consequently, we advocate for the implementation of a contrastive learning loss as a regulatory measure to alleviate the impact of multiple emotions and mitigate data imbalances. Contrastive learning is a training technique that originated from unsupervised learning. Supervised learning studies [7] have also demonstrated their effectiveness, utilizing samples from the same class as positive samples and others as negative samples. The loss used is the following:

$$L_{SupCon} = -\sum_{i \in I} \frac{1}{|P(i)|} \sum_{p \in P(i)} \log \frac{\exp(z_i \cdot z_p / \tau)}{\sum_{a \in A(i)} \exp(z_i \cdot z_a / \tau)}$$

Here, I is the set of classes, A(i) is the batch of samples contrasting with feature zi, P(i) is the set of positive samples of feature zi in A(i), i.e. samples with the same label.

The loss function is characterized by a vague description, suggesting that the feature extracted from a given sample should exhibit proximity to features extracted from positive samples while maintaining distance from features of other negative samples. Unlike traditional supervised learning, which prescribes a specific point in a lower dimension for a sample, contrastive learning defines positions in high-dimensional space that a sample should either approach or diverge from. This can mitigate the impact of labels, thereby diminishing the influence of multiple emotions.

As depicted in Fig. 3 and Fig. 1, the contrastive learning loss is computed from the feature projector's output, whereas the conventional cross-entropy loss relies on the output of the predictor. The feature projector and the predictor are both one-layer MLP. Therefore, the final loss can be represented as

$$L = \frac{L_{CE} + \alpha \cdot L_{SupCon}}{1 + \alpha}$$

where α is a hyperparameter to control the importance of contrastive learning loss in the final loss.

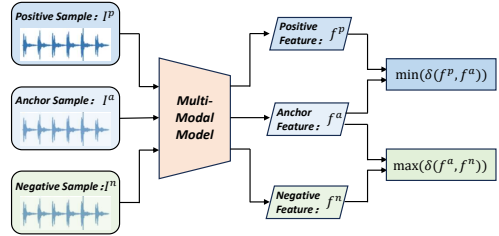

**Fig. 3.** The pipeline of the contrastive learning.

### 3.5 Data Augmentation

We formulate data augmentation strategies to mitigate the impact of noise, thereby improving the overall generalization of the model. In detail, we augment the audio signals in three ways: adding noise based on SNR, applying pitch shifts, and employing time stretching. When adding noise to the audio feature, we use an SNR of 30dB and randomly initialize the noise in Gaussian distribution. The pitch shift and time stretch are implemented by the librosa. In *IEMOCAP*, to increase the contrastive samples, we take advantage of the Dropout layers in our model. We run the prediction twice in one epoch to generate different features from the same sample. Also, as described in the previous section, we adopted MoCo [6] with size 16384.

## 4 Evaluation

### 4.1 Dataset

In addition to utilizing Chinese as the primary language, a key distinction between our dataset and existing ones is that we gather real-world data rather than employing actors to simulate various emotions. We collected 7,477 daily conversations from over 100 different people to create the dataset. For each sample, we hired several professional emotion analysis experts to analyze the data emotion and get artificial emotion classification labels (i.e., angry, fear, happy, neutral, sad, surprise) as ground truth. Fig. 4 presented the distribution of our dataset.

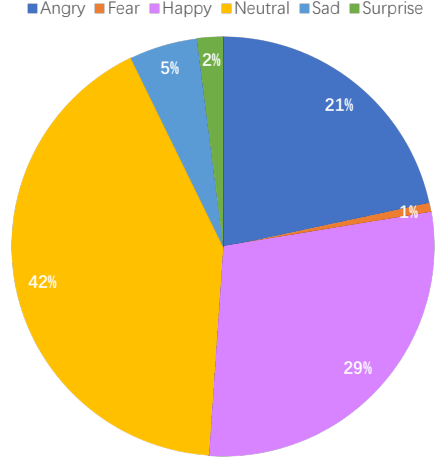

**Fig. 4.** Dataset distribution.

### 4.2 Experimental Setup

In this study, we implement a prototype of a multi-modal emotion recognition algorithm and evaluate the performance on the *VCEMO* dataset. A server equipped with 188GB RAM and a 48.0GB VRAM's NVIDIA TESLA A40 is used for the whole computation for the system.

**Training** We trained the models in both our proposed dataset *VCEMO* and another public dataset *IEMOCAP*. In both datasets, the model was trained for 50 steps with a batch size of 256. The optimizer used is Adam. Also, as a mostly used setting, the feature projector projects the feature into a 128-d vector. The temperature of contrastive learning loss t is 1. In *VCEMO*, we set learning rate to 1e−5, weight decay to 1e−3 and α to 0.1 while using 1e−4, 0 and 100 in *IEMOCAP*.

**Word Embeddings** We utilize a 300-dimensional GloVe [10] pre-trained embedding obtained from spaCy to encode the transcription into fixed-length vectors.

**Evaluation Metrics** Our dataset supports two kinds of setups. The first setup uses all samples for a 6-way classification, while the other setup only uses 4 classes. The 4 classes include angry, happy, neutral, and sad, which is a common setting for emotion recognition. We use two metrics to measure the performance of models comprehensively. The first metric is the accuracy of classification. Besides that, we adopt the F1-score as another metric to provide a more balanced evaluation of the model's performance. The F1-score is of the form:

$$F1 = \frac{2 \cdot (precision \cdot recall)}{precision + recall}$$

By considering both precision and recall, the F1-score can reflect the bias of model prediction to show whether a model achieves high accuracy by predicting those majority classes.

In *IEMOCAP*, we measured metrics following existing works. Weighted accuracy and unweighted accuracy are both considered to evaluate our model and existing works.

### 4.3 Micro Benchmark

Model Comparison In *VCEMO*, we split the dataset into 8/1/1 for train/val/test setting. We trained our model in the training part consisting 80% data of the dataset. The final model is chosen according to their performance on the 10% validation part. The test model is only used when the final model is determined. To show the effectiveness of our dataset and evaluate the performance of our model, we implement three other models as a comparison (i.e., Xu's model [20], UMONS [2] and Yoon's model [21]). The result of different models is shown in Table 1. Our model surpasses all other models in performance. In contrast to these models, our model can leverage external knowledge provided by BERT embeddings. Furthermore, our model enables improved multi-modal fusion through enhanced attention mechanisms.

In *IEMOCAP*, we followed the previous works using a 5-fold cross-validation. Each session in *IEMOCAP* will be used as a validation set once when training on the other 4 sessions. The final result is the average of all 5-fold results. Furthermore, when using contrastive learning regulation, our model is even better than CME[9] which requires addition alignment between transcription and acoustic signal.

Table 1. Comparison of different models on *VCEMO*.

| Models | 4-classes | | 6-classes | |
|---|---|---|---|---|
| | Accuracy | F1-score | Accuracy | F1-score |
| Ours (SupCon) | **67.40%** | **67.22%** | **65.51%** | **64.67%** |
| Ours | 66.99% | 66.90% | 65.37% | 64.54% |
| UMONS | 63.27% | 63.44% | 61.36% | 61.35% |
| Yoon's Model | 60.96% | 59.14% | 57.89% | 55.42% |
| Xu's Model | 59.42% | 57.68% | 58.16% | 55.39% |

Table 2. Results on *IEMOCAP* dataset.

| Metric | Xu's | CME | Ours | Ours (Sup-Con) |
|---|---|---|---|---|
| WA | 70.41% | 72.72% | 71.23% | **73.07%** |
| UA | 69.52% | 73.57% | 72.46% | **74.10%** |

### 4.4  Ablation Experiment

To further understand the effect of each modality, we performed an ablation study based on the 4-classes setup. The result is presented in Table 3.

**Impact of Transcription Modality** Theoretically, all the information presented in the text should also be contained within audio signals, suggesting that using only audio signals ought to outperform using only text modality. However, it's noteworthy that using only word embeddings outperformed using only one of the other two modalities. The disappointment with the result of using only acoustic signals may be due to that the information in audio signals is more challenging to extract, making it harder for the model to discern what is essential from the abundance of information. And using only BERT embeddings is slightly worse than using only acoustic signals. The reason is that encoding transcription with a pre-trained BERT model could cause a loss of information that is helpful in downstream tasks while trivial in upstream tasks. Therefore, word embeddings contain origin features and are easiest to extract, leading to a significant improvement in performance by around 4%. This gives us a hint that utilizing a text modality could help the model effectively extract features from the audio signals.

**Impact of BERT** When comparing experiments that only differ in the use of word embeddings or BERT embeddings, it's evident that using word embeddings outperforms using BERT embeddings in both single-modal and multimodal settings with the audio signal. This indicates that the knowledge within the database is still more important than external knowledge. However, adding BERT embeddings to word embeddings consistently improves performance by 1.5%, demonstrating that external knowledge can compensate for missing features from internal knowledge.

**Impact of contrastive learning regulation** In all benchmarks, additional contrastive learning regulation does improve the performance of our model. Especially in *IEMOCAP*, we can see it can significantly improve the performance of our model by over 1.5%. This is consistent with our expectation that contrastive learning regulation can reduce the effect of multi-label emotion recognition. Considering that *IEMOCAP* is using a much larger α, the result indirectly suggests that the ground truth of samples of *IEMOCAP* is more vague than our dataset *VCEMO*.

Table 3. Ablation study of using different modalities: Embeddings means the simple transcription embeddings while the BERT means the BERT embeddings.

| Used modality | Accuracy | F1-score |
|---|---|---|
| Embeddings | 57.40% | 55.05% |
| BERT | 53.29% | 50.16% |
| Acoustic | 53.18% | 51.15% |
| Embeddings + BERT | 57.26% | 55.11% |
| Embedding + Acoustic | 65.21% | 64.40% |
| BERT + Acoustic | 61.23% | 60.91% |
| Embeddings + BERT + Acoustic | **69.52%** | **74.10%** |

## 5  Conclusion

In this paper, we propose the emotion recognition dataset *VCEMO* for Chinese voiceprints. Compared with existing Chinese datasets, the proposed dataset is richer and more diversified in terms of voice tones and textual contents, containing more than 100 users and 7747 textual contents; the samples are all from daily conversations, which is closer to real-life scenarios. In addition, this paper proposes a multimodal emotion recognition model, which utilizes the co-attention structure for multimodal fusion. The contrastive-learning-based regulation training system achieves significantly better performance than SOTA on the *VCEMO* and *IEMOCAP* datasets.

## 6  Acknowledgement

This work is supported in part by the NSFC (61936015,62072306).

## References

1. Busso, C., Bulut, M., Lee, C.C., Kazemzadeh, A., Mower, E., Kim, S., Chang, J.N., Lee, S., Narayanan, S.S.: Iemocap: Interactive emotional dyadic motion capture database. Language resources and evaluation 42, 335-359 (2008)
2. Delbrouck, J.B., Tits, N., Brousmiche, M., Dupont, S.: A transformer-based joint-encoding for emotion recognition and sentiment analysis. ACL 2020 p. 1 (2020)
3. El Ayadi, M.M., Kamel, M.S., Karray, F.: Speech emotion recognition using Gaussian mixture vector autoregressive models. In: ICASSP. vol. 4, pp. IV-957. IEEE (2007)
4. Gat, I., Aronowitz, H., Zhu, W., Morais, E., Hoory, R.: Speaker normalization for self-supervised speech emotion recognition. In: ICASSP 2022. pp. 7342-7346. IEEE (2022)
5. Ghosh, S., Tyagi, U., Ramaneswaran, S., Srivastava, H., Manocha, D.: Mmer: Multimodal multi-task learning for speech emotion recognition. arXiv preprint arXiv:2203.16794 (2022)
6. He, K., Fan, H., Wu, Y., Xie, S., Girshick, R.:Momentumcontrastforunsupervised visual representation learning. In: CVPR. pp. 9729–9738 (2020)
7. Khosla, P., Teterwak, P., Wang, C., Sarna, A., Tian, Y., Isola, P., Maschinot, A., Liu, C., Krishnan, D.: Supervised contrastive learning. NeurIPS 33, 18661–18673 (2020)
8. Lee, C.C., Mower, E., Busso, C., Lee, S., Narayanan, S.: Emotion recognition using a hierarchical binary decision tree approach. Speech Communication 53(9-10), 1162–1171 (2011)
9. Li, H., Ding, W., Wu, Z., Liu, Z.: Learning fine-grained cross-modality excitement for speech emotion recognition. arXiv preprint arXiv:2010.12733 (2020)
10. Pennington, J., Socher, R., Manning, C.D.: Glove: Global vectors for word representation. In: EMNLP. pp. 1532–1543 (2014)
11. Radford, A., Kim, J.W., Xu, T., Brockman, G., McLeavey, C., Sutskever, I.:Robust speech recognition via large-scale weak supervision. In: ICML. pp. 28492–28518. PMLR (2023)
12. Schuller, B., Rigoll, G., Lang, M.: Hidden Markov model-based speech emotion recognition. In: ICASSP. vol. 2, pp. II–1. IEEE (2003)
13. Seehapoch, T., Wongthanavasu, S.: Speech emotion recognition using support vector machines. In: 2013 KST. pp. 86–91. IEEE (2013)

14. Stuhlsatz, A., Meyer, C., Eyben, F., Zielke, T., Meier, G., Schuller, B.: Deep neural networks for acoustic emotion recognition: Raising the benchmarks. In: ICASSP. pp. 5688–5691. IEEE (2011)
15. Triantafyllopoulos, A., Liu, S., Schuller, B.W.: Deep speaker conditioning for speech emotion recognition. In: ICME. pp. 1–6. IEEE (2021)
16. Trigeorgis, G., Ringeval, F., Brueckner, R., Marchi, E., Nicolaou, M.A., Schuller, B., Zafeiriou, S.: Adieu features? end-to-end speech emotion recognition using a deep convolutional recurrent network. In: ICASSP. pp. 5200–5204. IEEE (2016)
17. Vaswani, A., Shazeer, N., Parmar, N., Uszkoreit, J., Jones, L., Gomez, A.N., Kaiser, L., Polosukhin, I.: Attention is all you need. NeurIPS 30 (2017)
18. Wang, K., An, N., Li, B.N., Zhang, Y., Li, L.: Speech emotion recognition using Fourier parameters. IEEE Transactions on affective computing 6(1), 69–75 (2015)
19. Wang, Y., Boumadane, A., Heba, A.: A fine-tuned wav2vec 2.0/Hubert benchmark for speech emotion recognition, speaker verification and spoken language understanding. arXiv preprint arXiv:2111.02735 (2021)
20. Xu, H., Zhang, H., Han, K., Wang, Y., Peng, Y., Li, X.: Learning alignment for multimodal emotion recognition from speech. Proc. Interspeech 2019 pp. 3569–3573 (2019)
21. Yoon, S., Byun, S., Dey, S., Jung, K.: Speech emotion recognition using multi-hop attention mechanism. In: ICASSP 2019. pp. 2822–2826. IEEE (2019)
22. Yoon, S., Byun, S., Jung, K.: Multimodal speech emotion recognition using audio and text. In: 2018 SLT. pp. 112–118. IEEE (2018)
23. Yu, Z., Yu, J., Cui, Y., Tao, D., Tian, Q.: Deep modular co-attention networks for visual question answering. In: CVPR. pp. 6281–6290 (2019)
24. Zhao, M., Adib, F., Katabi, D.: Emotion recognition using wireless signals. In: Mobicom. pp. 95–108 (2016)